\documentclass[12pt]{article}
\usepackage{epsf,amsfonts,amssymb,epsfig,amsmath}
\renewcommand{\text}[1]{#1}

\newcommand{\be}{\begin{equation}}
\newcommand{\ee}{\end{equation}}
\newcommand{\ben}{\begin{displaymath}}
\newcommand{\een}{\end{displaymath}}
\newcommand{\bea}{\begin{eqnarray}}
\newcommand{\eea}{\end{eqnarray}}
\newcommand{\bean}{\begin{eqnarray*}}
\newcommand{\eean}{\end{eqnarray*}}
\newcommand{\nn}{\nonumber \\}
\newcommand{\ba}{\begin{array}}
\newcommand{\ea}{\end{array}}
\newcommand{\bi}{\begin{itemize}}
\newcommand{\ei}{\end{itemize}}

\newcommand{\reef}[1]{(\ref{#1})}

\def\l{\lambda}

\def\e{\epsilon}
\def\s{\sigma}
\def\e{\epsilon}
\def\r{\rho}

\def\m{\mu}
\def\n{\nu}

\newcommand{\bbR}{{\mathbb{R}}}

\newcommand{\rtw}{R^2_{(2)}}

\newcommand{\dub}{W}
\newcommand{\Z}{Z}

\newcommand{\dd}{\mathrm{d}}

\newcommand{\sss}{ {\sqrt{1-z}}}

\newcommand{\2}{{(2)}}

\def\oneone{\rlap 1\mkern4mu{\rm l}}

\allowdisplaybreaks  %Introduces page breaks inside \eqnarray

\begin{document}

\makeatletter
\renewcommand{\theequation}{\thesection.\arabic{equation}}
\@addtoreset{equation}{section}
\makeatother

\baselineskip 18pt

\begin{titlepage}

\vfill

\begin{flushright}
Imperial/TP/2007/JG/04\\
%hep-th/yymmnnn\\
\end{flushright}

\vfill

\begin{center}
   \baselineskip=16pt
   {\Large\bf $D=5$ $SU(2)\times U(1)$ Gauged Supergravity\\[8pt] from $D=11$ Supergravity}
   \vskip 2cm
      Jerome P. Gauntlett and Oscar Varela\\
   \vskip .6cm
      \begin{small}
      \textit{Theoretical Physics Group, Blackett Laboratory, \\
        Imperial College London, London SW7 2AZ, U.K.}
        %E-mail: j.gauntlett, d.waldram@imperial.ac.uk}
        \end{small}\\*[.6cm]
      \begin{small}
      \textit{The Institute for Mathematical Sciences, \\
        Imperial College London, London SW7 2PE, U.K.}
        %E-mail: j.gauntlett, d.waldram@imperial.ac.uk}
        \end{small}
   \end{center}

\vfill

\begin{center}
\textbf{Abstract}
\end{center}

\begin{quote}
We consider the most general class of supersymmetric solutions of
$D=11$ supergravity consisting of a warped product of $AdS_5$ with a
six-dimensional internal manifold ${\cal N}_6$, which are dual to
$N=2$ super conformal field theories in $d=4$. For any such ${\cal
N}_6$ we construct the full non-linear Kaluza-Klein ansatz for the
reduction of $D=11$ supergravity on ${\cal N}_6$ down to $D=5$
$SU(2)\times U(1)$ gauged supergravity, at the level of the bosonic
fields. This allows one to uplift any solution of the $D=5$
supergravity to obtain a solution of $D=11$ supergravity for any
given ${\cal N}_6$. Using an explicit ${\cal N}_6$, corresponding to
M5-branes wrapping holomorphic curves in a Calabi-Yau two-fold, we
uplift some solutions and comment upon their interpretation.

\end{quote}

\vfill

\end{titlepage}
\setcounter{equation}{0}

%%%%%%%%%%%%%%%%%%%%%%%%%%%%%%%%%%%%%%%%%%%%%%%%%%%%%%%%%%%%%%%%%%%%%%%
%\tableofcontents
%%%%%%%%%%%%%%%%%%%%%%%%%%%%%

\section{Introduction}
A Kaluza-Klein (KK) reduction of a higher dimensional theory of
gravity down to a lower dimensional theory using an internal manifold ${\cal N}$
is said to be consistent if any solution to the equations of motion of
the lower-dimensional theory can be uplifted on ${\cal N}$ to obtain
a solution to the equations of motion of the higher-dimensional
theory. If one keeps the entire infinite tower of KK modes the
reduction is obviously consistent, since the reduction is simply a rewriting of the
original higher dimensional theory. However, in certain special cases
it is possible to obtain consistent KK reductions by further truncating to a
finite number of modes.

The standard way to prove that a KK reduction is consistent is to
construct a KK ansatz: i.e. an explicit embedding of the fields of
the lower-dimensional theory into the higher-dimensional one, with
the property that the equations of motion of the higher-dimensional
theory are satisfied provided that the equations of motion of the
lower-dimensional theory are. Such a KK ansatz has the virtue that
any explicit solution of the lower-dimensional theory can be
uplifted to obtain an explicit solution of the higher dimensional
theory. This has proved to be a very powerful technique for
constructing supergravity solutions relevant for string/M-theory.

While much is now known about consistent KK truncations
%(see e.g.
%\jpg{fix} \cite{Tsikas:1986rx,Duff:1984hn,Pope:1985jg,Buchel:2006gb}),
a single overarching principle governing all cases remains elusive, if
indeed one exists. Recently we put forward a conjecture \cite{gv} (related to
\cite{Duff:1985jd}) for sufficient criteria for consistency,
covering a large number of supersymmetric cases.
Consider the most general supersymmetric solutions of $D=10$
or $D=11$ supergravity\footnote{One can also consider supergravity
theories in other dimensions.} that are (warped) products,
$AdS_{d+1}\times_w {\cal N}$, of a $d+1$-dimensional anti-de-Sitter
space, $AdS_{d+1}$, with an internal space, ${\cal N}$, that are
dual to supersymmetric conformal field theories (SCFTs) in $d$
dimensions. We conjectured that there should be a consistent KK
reduction on ${\cal N}$ to a gauged supergravity theory in $d+1$
dimensions for which the fields are dual to those in the
superconformal current multiplet of the $d$-dimensional dual SCFT.

We now know that this conjecture is in fact a theorem for a number
of different cases for which the explicit KK reduction ansatze have
  been constructed. For example, for the class of $AdS_5 \times M_5$ solutions of type IIB supergravity,
where $M_5$ is a Sasaki-Einstein manifold, that are dual to $N=1$ SCFTs in $d=4$,
it was shown in \cite{Buchel:2006gb} (see also \cite{Tsikas:1986rx}) that there is a consistent KK reduction of IIB supergravity on $M_5$ to minimal
gauged supergravity in $D=5$.
This result was generalised in \cite{gv} to the most general class of $AdS_5\times_w {\cal N}_5$ solutions
of type IIB that are dual to $N=1$ SCFTs in $d=4$ \cite{Gauntlett:2005ww}, thus verifying
the conjecture in this case.
It was shown in \cite{gov} that minimal $D=5$ gauged supergravity also arises from the KK reduction of $D=11$ supergravity
on ${\cal N}_6$ associated with the most general class of $AdS_5 \times_w {\cal N}_6$ solutions
of $D=11$ with $d=4$, $N=1$ SCFT duals \cite{gmsw}.
Furthermore, \cite{gov} also showed that for general classes of $AdS_4\times_w{\cal N}_7$ solutions that are dual
to $N=2$ SCFTs in $d=3$ there is a consistent KK reduction on ${\cal N}_7$ to $N=2$ gauged supergravity in $D=4$.
Of course, the well-known consistent KK reductions of $D=11$ supergravity on $S^4$  \cite{deWit:1986iy} and
$S^7$ \cite{Nastase:1999cb,Nastase:1999kf}, or of IIB supergravity\footnote{No
complete reduction ansatz of IIB supergravity on $S^5$ has been yet constructed.}
on $S^5$ (see, {\it e.g.}, \cite{Cvetic:1999xp,Cvetic:2000nc,Khavaev:1998fb,Lu:1999bw}),
related to the maximally supersymmetric solutions $AdS_7 \times S^4$,
$AdS_4 \times S^7$ and $AdS_5 \times S^5$, respectively, are also examples supporting the conjecture.

In this paper we will consider the general class of $AdS_5\times_w {\cal N}_6$
solutions of $D=11$ supergravity that are dual to $N=2$ SCFTs in
$d=4$. Such supergravity solutions were classified by Lin, Lunin and
Maldacena in \cite{llm}, refining the work of \cite{gmsw}. Such
SCFTs have an $SU(2)\times U(1)$ R-symmetry and so the conjecture of
\cite{gv} says that there should be a consistent KK reduction of
$D=11$ supergravity on ${\cal N}_6$ to Romans' $D=5$ $SU(2)\times
U(1)$ gauged supergravity \cite{rom} (more precisely to what is
called the $N=4^+$ theory in \cite{rom}). In this paper we will
construct the full non-linear KK ansatz for the bosonic fields.

At a technical level, this case is considerably more involved than
the previous cases considered in \cite{gov,gv}. The central subtlety
in guessing the correct KK ansatz is the correct incorporation of
the scalar field of the $D=5$ gauged supergravity. We found the
results of \cite{Lu:1999bw,Cvetic:2000yp} to be particularly
helpful. In \cite{Lu:1999bw} the full KK ansatz for the
reduction of type IIB supergravity on an $S^5$ to Romans' theory was
presented. This is expected to be a truncation of a more general
KK ansatz to maximally supersymmetric $SO(6)$ gauged supergravity. Now,
after T-duality and uplifting, it is known \cite{Cvetic:2000cj} that
the $AdS_5\times S^5$ solution of type IIB supergravity can be used
to obtain the singular $AdS_5\times_w{\cal N}_6$ solution of $D=11$
supergravity found in \cite{Alishahiha:1999ds}. By performing the
same T-duality and uplifting on the type IIB KK reduction ansatz for
the $S^5$ found in \cite{Lu:1999bw}, it was shown in
\cite{Cvetic:2000yp} how one can obtain Romans' theory by reduction
of the specific $D=11$ solution found in \cite{Alishahiha:1999ds}.
The form of the KK reduction ansatz, for this specific solution,
provided us with important clues in obtaining the ansatz for an
arbitrary $AdS_5\times_w{\cal N}_6$ solution that we shall present
here.

The only regular $AdS_5\times_w {\cal N}_6$ solution that we are
aware of is the solution constructed by Maldacena and N\'u\~nez in
\cite{mn}. This solution is dual to the $N=2$ SCFT in $d=4$ that
lives on $M5$-branes wrapping holomorphic Riemann surfaces in
Calabi-Yau two-folds. Using this solution we can uplift any explicit
solution of Romans' theory to obtain an explicit solution of $D=11$
supergravity. We uplift some known solutions of Romans' theory and
discuss how some of them are related to wrapped brane solutions.

The calculations required for checking that our KK ansatz is
correct are quite involved and so we have included a few details
in an appendix.

\section{Romans from Lin, Lunin and Maldacena via Ka\-lu\-za and Klein}
\label{RLLMKK}

In this section we present the KK ansatz for the reduction of
$D=11$ supergravity on the geometries ${\cal N}_6$ classified by
Lin, Lunin and Maldacena (LLM), down to Romans' $D=5$ $SU(2) \times U(1)$ gauged supergravity.
We begin by first reviewing the work of LLM and Romans in
subsections \ref{undefgeom} and \ref{secRomans}, respectively.

\subsection{The geometry of LLM} \label{undefgeom}

The geometry underlying the most general $AdS_5$ solutions of $D=11$
supergravity that are dual to $N=2$ SCFTs in $d=4$ was first derived
by LLM in \cite{llm}, where it was shown
that such supergravity solutions are determined by solutions of a
continuous three dimensional version of the Toda equation. The same
conditions were rederived from a different point of view in
\cite{wb}, by taking the $AdS$ limit of a general class of
Minkowski geometries corresponding to M5-branes wrapped on a K\"ahler two-cycle in a
Calabi-Yau two-fold. We will use the notation of \cite{wb}, which
also includes the explicit dictionary between the two descriptions.
Our conventions for $D=11$ supergravity are as in
\cite{Gauntlett:2002fz}, some of which is recorded in appendix
\ref{KKdetails}.

The metric is a warped product of $AdS_5$, with radius $1/m$, with a
six-dimensional internal manifold ${\cal N}_6$:
\begin{equation} \label{undef11metric}
\dd s^2_{11} = \lambda^{-1} \dd s^2(AdS_5) + \dd s^2 ({\cal N}_6),
\end{equation}
where the warp factor $\lambda$ is a function of the coordinates on
${\cal N}_6$ only. As in \cite{wb}, we will let $(e^1,\dots,e^6)$ be
an orthonormal frame for ${\mathcal N}_6$,
\begin{equation}
\dd s^2 ({\cal N}_6) = (e^1)^2 + (e^2)^2 +(e^3)^2 +(e^4)^2 +(e^5)^2
+(e^6)^2 \; ,
\end{equation}
with
 \bea e^4=\frac{\l}{2m\sss}\dd \r\nn (e^5)^2 +
(e^6)^2=\frac{\lambda^2\rho^2}{4m^2} \dd\mu^i\dd\mu^i \; .\eea
Here, $\rho$ is a coordinate, $z\equiv \lambda^3\rho^2$ and the
$\mu^i$, $i=1,2,3$, satisfying $\mu^i\mu^i=1$, parametrise a
two-sphere. Note that the one-form $\hat\rho$ in \cite{wb} is
denoted here by $e^4$. We will define a positive orientation by
$\epsilon = e^{123456}$.
The four-form flux is given by
\bea\label{6.9} G_4=-\frac{1}{8m^2}\epsilon_{ijk}\mu^i
\dd\mu^j\wedge \dd\mu^k\wedge\left[
   \dd\left(\l^{1/2}\r\sss e^3\right)
   + 2m\left(\l\r e^{12}+\l^{-1/2}e^{34}\right)
   \right].
\eea
The necessary and sufficient conditions for (\ref{undef11metric}),
(\ref{6.9}) to be a supersymmetric solution to the equations of
motion of $D=11$ supergravity are
\begin{equation}
\label{65}
\begin{aligned}
   \dd\left(\l^{-1}\sss e^1\right) &=
      m\l^{-1/2}\left(\l^{3/2}\rho e^{14}+e^{23}\right), \\
   \dd\left(\l^{-1}\sss e^2\right) &=
      m\l^{-1/2}\left(\l^{3/2}\rho e^{24}-e^{13}\right), \\
   \dd\left(\frac{\l^{1/2}}{\sss}e^3\right) &=
      - \frac{2m\l}{1-z}e^{12}
      - \frac{3\l\rho}{(1-z)^{3/2}}\left[
         (\dd\l)_4 e^{12}
         - (\dd\l)_2 e^{14}
         + (\dd\l)_1 e^{24}\right],
%\star_6F&=&\l^{5/2}\Big(d[\l^{-5/2}\sss
%e^3]-4m[\l^{-2}e^{12}\nn&&+\l^{-1/2}\r e^{34}]\Big).
\end{aligned}
\end{equation}
where
 \be\label{dl} \dd \lambda=(\dd \lambda)_1 e^1+(\dd \lambda)_2
e^2+(\dd \lambda)_4 e^4 \ , \ee
and $e^{12} \equiv e^1 \wedge e^2$, etc.

The $d=4$, $N=2$ dual SCFTs have an $SU(2)\times U(1)$ R-symmetry,
and this manifests itself as isometries of the
internal metric $\dd s^2({\cal N}_6)$. The $SU(2)$ symmetry of the
two-sphere, parametrised by the $\mu^i$, is clearly a symmetry of
the solution. The vector field dual to $e^3$ is proportional to the
additional $U(1)$ Killing vector, consistent with \reef{dl}. This is
explained in more detail in \cite{wb} where it is shown how the
above conditions allow one to introduce coordinates, used by
\cite{llm}, which makes the $U(1)$ symmetry manifest. One subtlety
is that the frame we will be using depends on the coordinate
parametrising the orbits of this $U(1)$ and so, for example, we have
from \reef{65} \bea
{}({\dd(\dd\lambda)_1})_3&=&-\frac{m\lambda^{1/2}}{\sqrt{1-z}}(\dd\lambda)_2
\; ,\nn
{}({\dd(\dd\lambda)_2})_{3}&=&\frac{m\lambda^{1/2}}{\sqrt{1-z}}(\dd\lambda)_1
\; . \eea

\subsection{Romans' $D=5$ $SU(2)\times U(1)$ supergravity} \label{secRomans}

The field content of Romans' ($N=4^+$) $SU(2) \times U(1)$ gauged
supergravity in $D=5$ \cite{rom} consists of a metric, with line
element $\dd s^2_5$, a scalar field $X$, $U(1) \times SU(2)$ gauge
fields $B$, $A^i$, with $i=1,2,3$, and a complex two form $C$ which
is charged with respect to the $U(1)$ gauge field. The corresponding
field strengths for these potentials are given by
\bea \label{Bianchis} G&=&\dd B ,\nn F^i&=&\dd A^i-\tfrac{1}{\sqrt
2}m\epsilon_{ijk}A^j\wedge A^k , \nn F&=& \dd C+imB\wedge C . \eea
In our conventions, the equations of motion for the scalar and the
gauge fields are
%%%%%%%%%%%%%%%%%%%%%
\bea \label{eomromans1} \dd (X^{-1}\, {*\dd X}) &=&\tfrac{1}{3}
X^4\, {*G}\wedge G - \tfrac{1}{6} X^{-2} \, ({*F^i}\wedge F^i +
{*{C}}\wedge \bar C)\nn &&- \tfrac{4}{3}m^2\, (X^2 - X^{-1})\,
{*\oneone} ,\\
\label{eomromans2} \dd (X^4\, {*G}) &=& - \tfrac{1}{2} F^i\wedge F^i
-
\tfrac{1}{2} {\bar C}\wedge C ,\\
\label{eomromans3} D(X^{-2}\, {*F^i})&=&- F^i\wedge G,\\
 X^{2}\, {*F} &=&  i\, m\, C \,, \label{eomromans4} \eea
where $D(X^{-2}*F^i)\equiv \dd (X^{-2}*F^i)+{\sqrt
2}m\epsilon_{ijk}A^k\wedge (X^{-2}*F^j)$, we have taken
$\epsilon_{01234}=+1$ for the five-dimensional space, and $\bar C$
denotes complex conjugate of $C$. In addition, the Einstein equation
reads
\bea R_{\m\n} &=& 3 X^{-2}\,  \partial_\m X\, \partial_\n X -
\tfrac{4}{3}m^2\,(X^2 + 2  X^{-1})\, g_{\m\n}\nn & & + \tfrac{1}{2}
X^4 \, (G_\m{}^\r G_{\n \r} -\tfrac{1}{6} g_{\m\n} \,
G_{\r\s}G^{\r\s}) + \tfrac{1}{2} X^{-2}\, (F^{i\ \r}_\m F^{i}_{\n\r}
- \tfrac{1}{6} g_{\m\n}\, F^i_{\r\s}F^{i\r\s})\nn & & + \tfrac{1}{2}
X^{-2}\,  ({C}_{(\m}{}^\r\,  \bar C_{\n)\r} - \tfrac{1}{6}
g_{\m\n}\, C_{\r\s}\bar C^{\r\s})\, . \label{eomromans5} \eea
%%%%%%%%%%%%%%%%%%%%%%%%%%%%%%%%%

These equations of motion can be derived from the five-dimensional
Lagrangian given by
%%%%%%%%%%%%%%%
\bea {\cal L} &=& R\, {*\oneone} - 3 X^{-2} {* \dd X}\wedge \dd X
-\tfrac{1}{2} X^4\,  {*G}\wedge G -\tfrac{1}{2} X^{-2}\,
({*F^i}\wedge F^i + *C_\2\wedge \bar C)\nn & & -\frac{i}{2m} C\wedge
\bar F- \tfrac{1}{2} F^i\wedge F^i\wedge B + 4m^2(X^2 + 2 X^{-1})\,
{*\oneone}\, . \label{laromans} \eea
%%%%%
Note that the scalar $X$ can be written in terms of a
canonically-normalised dilaton $\phi$ as $X=e^{-\tfrac{1}{\sqrt6}\,
\phi}$. %These are precisely the equations of motion and Lagrangian
%for Romans' $D=5$ $SU(2)\times U(1)$ gauged supergravity theory. (We
This Lagrangian can be obtained from the one in \cite{rom}, up to an
overall factor, after changing the signature of the metric, taking
$g_1=-2m$, $g_2=-2{\sqrt 2}m$, $\xi=X^{-1}$ and scaling the gauge
fields by a factor of 1/2. We also note that if we set $m=-g$ we
have exactly the same equations of motion and Lagrangian as given in
\cite{Lu:1999bw}, except that we disagree with the definition of
$D(X^{-2}*F^i)$ by a sign.

Finally, for later use, we note that if we restrict to
configurations with $X=1$, $F^1=F^2=C=0$ and  $F^3={\sqrt 2} G$, the
equations of motion (\ref{eomromans1})--(\ref{eomromans5}) of Romans
theory truncate to
\begin{eqnarray}
&& \label{D=5eomFmin} \dd * G  = -G \wedge G  \\
&& \label{D=5Einmin} R_{\mu \nu} = -4 m^2 g_{\mu \nu} +\tfrac{3}{2}
( G_{\mu \lambda} G_\nu{}^\lambda -\tfrac{1}{6}  g_{\mu \nu}
G_{\lambda \rho} G^{\lambda \rho} )  .
\end{eqnarray}
These are the equations of motion of minimal $D=5$ gauged
supergravity \cite{Gunaydin:1983bi}. In particular, we can use the
reduction formulae given in subsection \ref{secKKred} to uplift any
solution of $D=5$ minimal gauged supergravity %on an ${\cal N}_6$
to obtain a solution of $D=11$ supergravity.

\subsection{The Kaluza-Klein ansatz} \label{secKKred}

We are now in a position to construct the full non-linear ansatz
for the KK reduction of $D=11$ supergravity on any of the
six-dimensional manifolds ${\cal N}_6$ reviewed in subsection
\ref{undefgeom} down to Romans' $D=5$ $SU(2)\times U(1)$ gauged
supergravity.

The KK ansatz for the metric takes the form
\begin{equation}\label{defmet}
\dd s^2_{11} = X^{-1/3} \Delta^{1/3} \l^{-1}\dd s^2_5 + \dd s^2
(\hat {\cal N}_6) \; ,
\end{equation}
where we have introduced the ubiquitous quantity
\begin{equation}
\Delta = X z + X^{-2}(1-z) \; .
\end{equation}
In addition
{\setlength\arraycolsep{0pt} \bea
   \dd && s^2(\hat{\cal N}_6) =
      X^{2/3} \Delta^{1/3}\! \left[ (e^1)^2 \! + \! (e^2)^2 \! +\!
      (e^4)^2\right] \!
      +\! X^{5/3} \Delta^{-2/3}(\hat e^3)^2 \!+ \!
      X^{-4/3} \Delta^{-2/3}\frac{\lambda^2\rho^2}{4m^2}
      D\mu^iD\mu^i \nonumber \\ &&
%&=&   e^1\otimes e^1 + e^2\otimes e^2 + e^3\otimes e^3  +  \frac{\lambda^2}{4m^2}\frac{\dd\rho^2}{1-z}
%            + \frac{\lambda^2\rho^2}{4m^2} \dd\mu^i\dd\mu^i
\eea }
where
\begin{eqnarray} \label{shift}
&& \hat e^3 = e^3 + \frac{\sss}{\l^{1/2}}B \ , \\
&& D\mu^i=\dd \mu^i+{\sqrt 2}m\epsilon_{ijk}A^k\mu^j \ .
\end{eqnarray}
The way that the $SU(2)\times U(1)$ gauge fields are incorporated
here follows the usual general principles of KK reductions. The way
that the scalar field enters is much less obvious as is the
expression for the four-form which is given by
\begin{eqnarray} \label{ansG}
G_4 = \tilde{{G_4}} +G \wedge \beta_2 + F^i \wedge \beta_2^i + *F^i
\wedge  \beta_{1}^i+ \left(C \wedge  \alpha_2 + F \wedge \alpha_1 +
c.c.\right)
\end{eqnarray}
where here and in the following $*$ is the Hodge dual with respect
to the metric $\dd s^2_5$, $c.c.$ denotes complex conjugation,
\bea \tilde{{G_4}}&=&-\frac{1}{8m^2}\epsilon_{ijk}\mu^iD\mu^j\wedge
D\mu^k\wedge\bigg[ \dd\left(X^{-2}\Delta^{-1}\rho(1-z)\right) \wedge
\frac{\l^{1/2}}{\sss}\hat e^3\nn
&+&X^{-2}\Delta^{-1}\rho(1-z)\dd\left(\frac{\l^{1/2}}{\sss}e^3\right)
   +2m\left(\l\r e^{12}+\l^{-1/2}\hat e^{34}\right)
   \bigg] \label{ansGtilde}
\eea
(we emphasise that in the second term there is no hat on the $e^3$
--the term should be constructed from \reef{65}), $\hat{e}^{34}
\equiv \hat{e}^3 \wedge e^4$ (and, in general, $\hat{e}^{3 a_1
\cdots a_k} \equiv \hat{e}^3 \wedge e^{a_1 \cdots a_k}$)  and
\bea \beta_ 2 &=& \frac{1}{8m^2}\rho z
X\Delta^{-1}\epsilon_{ijk}\mu^iD\mu^j\wedge D\mu^k \ , \nn
\beta_{2}^i &=& \frac{1}{2{\sqrt 2}m} \left[ X^{-2}\Delta^{-1}\rho
\lambda^{1/2}\sss D\mu^i\wedge \hat e^3 -2m \mu_i (\lambda \rho
e^{12} + \lambda^{-1/2} \hat e^{34}) \right] \ , \nn \beta_1^i &=&
-\frac{X^{-2}}{2{\sqrt 2}m}\left( \mu^i \dd \rho +\rho D\mu^i\right)
\ , \nn \alpha_1&=&\frac{1}{2{\sqrt 2}m}\l^{-1}\sss(e^1-ie^2) \ ,\nn
\alpha_2&=&\frac{1}{2{\sqrt 2}}(e^1-ie^2)\wedge \left(\l\r
e^4+i\l^{-1/2}\hat e^3\right) \ .  \eea
Of course, when the $D=5$ fields are trivial, $X=1$,
$B=0$, $A^i=0$, $C=0$, the KK reduction ansatz (\ref{defmet}),
(\ref{ansG}) reduces to the undeformed geometry
(\ref{undef11metric}), (\ref{6.9}).

After a lengthy calculation, one can show that the KK reduction
ansatz (\ref{defmet}), (\ref{ansG}) satisfies the equations of
motion of $D=11$ supergravity (\ref{Bianchi11})--(\ref{Einstein11}),
provided the $D=5$ fields satisfy the equations of motion
(\ref{eomromans1})--(\ref{eomromans5}) of Romans theory. This shows,
at the level of the bosonic fields, the consistency of the KK
reduction. See appendix \ref{KKdetails} for some details of the
consistency proof.

\section{Uplifting explicit solutions and wrapped branes}

The only regular $AdS_5\times_w {\cal N}_6$ solution with $N=2$
supersymmetry that we are aware of is the solution found by
Maldacena and N\'u\~nez in \cite{mn}. This solution is dual to the
$N=2$ $d=4$ CFT living on M5-branes wrapping a Riemann surface
holomorphically embedded in a Calabi-Yau two-fold. More precisely,
the CFT is obtained after taking a decoupling limit and then flowing
to the far IR as we will elaborate on a little further below. Using
the explicit formulae of subsection \ref{secKKred}, this solution
can be used to uplift explicit solutions of Romans' $D=5$
$SU(2)\times U(1)$ gauged supergravity to obtain explicit solutions
of $D=11$ supergravity.

Setting $m=1/2$ for simplicity, we find that the $D=11$ metric
(\ref{defmet}) of the uplifted solution takes now
the explicit form
\bea\label{ul} \dd s^2_{11} &=& 2^{-2/3}\bar\Delta^{1/3} \dd
s^2_5+2^{1/3}X\bar\Delta^{1/3}[\dd  \theta^2+ \dd  s^2(H_2)]\nn
&+&2^{1/3}X\bar\Delta^{-2/3}\sin^2\theta(\dd x_3+V+\tfrac{1}{2}B)^2
+2^{-2/3}X^{-2}\bar\Delta^{-2/3}\cos^2\theta D\mu^iD\mu^i \eea
where we have replaced, for convenience, the
coordinate $\rho$ by a new coordinate $\theta$ such that $\rho =
\tfrac{1}{2} \cos \theta$, $x_3$ is a coordinate along the $U(1)$
Killing direction $e^3$, $ds^2(H_2)$ is the standard metric on a
unit radius hyperbolic two plane, $dV=-vol(H_2)$ and
$\bar\Delta=\cos^2\theta+(X^{-3}/2)\sin^2\theta$. We note that here
and in the following we can quotient $H_2$ by a discrete group of
isometries, $H_2/\Gamma$, and hence obtain a compact Riemann surface
with genus greater than one, without breaking supersymmetry. This is
the Riemann surface that the M5-branes are wrapping. In obtaining
\reef{ul} we have used the expression for the solution of \cite{mn}
as given in \cite{llm} and then used appendix D of \cite{wb} to
translate it into the language of this paper. For example, we note
that $z=2\cos^2\theta/(1+\cos^2\theta)$.

In the following subsections, we will use \reef{ul} to uplift some
explicit supersymmetric solutions of Romans' $D=5$ theory to obtain
explicit
solutions of $D=11$ supergravity, some of which are new. % The discussion will naturally
%lead to dome discussion of some related sueprgravity solutions both
%in $D=11$ and in type IIB supergravity.
We will also discuss how these solutions are related to other
solutions in $D=11$ and IIB supergravity.

\subsection{Uplifting the Nieder-Oz solution}
Following \cite{mn}, Nieder and Oz considered the following ansatz
for the $D=5$ supergravity fields \cite{no} (translated into our
conventions):
\bea\label{noflow}
ds^2_5&=&\frac{e^{2f}}{4m^2}\left[-dt^2+dr^2\right]+
\frac{e^{2g}}{4m^2}ds^2(H_3)\nn X&=&e^{-\varphi}\nn
A^i&=&\frac{1}{2{\sqrt 2}m}\epsilon_{ijk}\omega^{jk}\nn B&=&0 \; ,
\quad C=0 \eea
where $\omega^{ij}$ is the spin connection of the unit radius metric
on a three-dimensional hyperboloid $H_3$ and
$f,g,\varphi$ are functions of $r$ only. Once again, here and in the
following, one can replace $H_3$ with $H_3/\Gamma$, whilst
preserving supersymmetry. This is indeed a supersymmetric solution
providing that $f,g,\varphi$ satisfy
\bea\label{nozode}
e^{-f}\dot
f&=&-\frac{1}{3}e^{-\varphi}-\frac{1}{6}e^{2\varphi}-e^{\varphi-2g}\nn
e^{-f}\dot
g&=&-\frac{1}{3}e^{-\varphi}-\frac{1}{6}e^{2\varphi}+e^{\varphi-2g}\nn
e^{-f}\dot
\varphi&=&-\frac{1}{3}e^{-\varphi}+\frac{1}{3}e^{2\varphi}-e^{\varphi-2g}
\eea
In \cite{no} these equations were partially integrated.
Furthermore, it was shown numerically that there are solutions that
interpolate from a region where the $D=5$ metric is
\be\label{ar} ds^2_5=\frac{1}{m^2r^2}[-dt^2+ds^2(H_3)+dr^2] \ee
down to an exact $AdS_2\times H_3$ solution given by \bea\label{gss}
ds^2_5&=&\frac{1}{4m^22^{2/3}}\left[ds^2(AdS_2)+ 4
ds^2(H_3)\right]\nn X&=&4^{-1/3}\nn A^i&=&\frac{1}{2{\sqrt
2}m}\epsilon_{ijk}\omega^{jk}\eea and $B=0$, $C=0$. Such solutions
are sometimes called topological black holes.

In \cite{no} these solutions were uplifted on an $S^5$ using the results
of \cite{Lu:1999bw} to obtain solutions of type IIB supergravity\footnote{The correctly uplifted formula, for the $AdS_2\times H_3$ solution,
were given in \cite{d3}.}. The type IIB solution obtained by uplifting the $AdS_2\times H_3$ solution \reef{gss} is dual to the SCFT
living on D3-branes wrapping an associative $H_3$ in a $G_2$ manifold.
This CFT preserves 2 supercharges. A key aspect of this interpretation is that the gauge fields that
are switched on are dual to the $R$-symmetry currents that must be activated in order that the field theory on the D3-branes,
i.e. $N=4$ $d=4$ SYM theory on $\bbR\times H_3$, can preserve supersymmetry \cite{mn}.
Additional evidence for this interpretation is provided by the
uplifted numerical solution of \cite{no}. It describes a ``RG flow across dimensions'' from the locally asymptotic
$AdS_5$ region \reef{ar}, where one clearly sees the $\bbR\times H_3$ world-volume of the D3-brane,
down to the $AdS_2$ fixed point \reef{gss}. Exactly the same kind of arguments \cite{mn} lead to the interpretation of
the $AdS_5\times_w {\cal N}_6$ solution with $N=2$ supersymmetry that we mentioned
at the beginning of this section.

After setting $m=1/2$ we can use \reef{ul} to uplift the
$AdS_2\times H_3$ solution \reef{gss} and also the numerical
solution found in \cite{no} to obtain solutions of $D=11$
supergravity. We find that the $AdS_2\times H_3$ solution \reef{gss}
uplifts to a $D=11$ solution with metric given by
\bea\label{ul2} ds^2_{11} &=&
\frac{(1+\sin^2\theta)^{1/3}}{2^{4/3}}\bigg[ds^2(AdS_2)+4ds^2(H_3)+2ds^2(H_2)\nn
&+&2d\theta^2+\frac{2\sin^2\theta}{(1+\sin^2\theta)}(dx_3+V)^2
+\frac{4\cos^2\theta}{(1+\sin^2\theta)}D\mu^iD\mu^i\bigg] \eea
where $D\mu^i=d\mu^i+\omega^{ij}\mu^j$.
The numerical solution of \cite{no} uplifts to a $D=11$ solution that interpolates from a
region with a locally $AdS_5$ factor as in \reef{ar}, to the above $AdS_2$
solution. The dual interpretation is therefore clear. The numerical solution describes the
RG flow across dimensions from the $d=4$ CFT that lives on M5-branes
wrapping a holomorphic $H_2$ in a $CY_2$, placed on $\bbR\times H_3$ with suitable $R$
symmetry currents activated in order to preserve supersymmetry, down to a
conformal quantum mechanics with two supercharges.

One might wonder what happens if one considers the CFT living on
M5-branes placed directly on $\bbR\times H_3\times H_2$. With the
above $R$-symmetry currents, this corresponds to $M5$-branes
wrapping $H_3\times H_2$ in a $CY_3\times CY_2$ where $H_3$ is a
SLAG 3-cycle and $H_2$ is a holomorphic 2-cycle. In fact solutions
describing such wrappings were already constructed in \cite{gk}. In
particular, there is an $AdS_2$ solution which is exactly the same as
the uplifted solution \reef{ul2}. Furthermore, \cite{gk} also
studied some flow equations: if we substitute
$e^{2f}=2^{2/3}e^{-4\lambda/3}e^{2\bar f}$,
$e^{2g}=2^{5/3}e^{-4\lambda/3}e^{2g_1}$ and
$e^{-\varphi}=2^{-1/3}e^{-10\lambda/3}$ into \reef{nozode} then we
obtain the odes in equation (4.5) of \cite{gk} provided that we set $g_2=-\lambda$.

We can also consider reversing the order. We could first consider the $d=3$ CFT living
on M5-branes wrapping a SLAG $H_3$ in a $CY_3$. The relevant $AdS_4\times_w {\cal N}_7$
solution was constructed in \cite{gkw}. We now consider placing
this $d=4$ CFT on $\bbR\times H_2$ with suitable $R$-symmetry currents to preserve supersymmetry
and ask what happens in the IR. It was shown in \cite{gv} how one can carry out
an explicit KK reduction from $D=11$ supergravity on ${\cal N}_7$
to $D=4$ minimal gauged supergravity.
In particular, one finds that the $AdS_2\times H_2$ solution of \cite{ck} also leads to
the solution \reef{ul2}. Furthermore, the more general explicit topological black hole
solution \cite{ck}  uplifted on ${\cal N}_7$ describes the flow from the three-dimensional
CFT on $\bbR\times H_2$ to the conformal quantum mechanics.

\subsection{Uplifting a Maldacena-N\'u\~nez solution}
Let us now consider another class of wrapped brane solutions. In
\cite{mn} Maldacena and N\'u\~nez constructed supersymmetric
solutions of $D=5$ $U(1)^3$ gauged supergravity which, upon lifting
on an $S^5$ to type IIB supergravity, describe the $(2,2)$ SCFT
arising on $D3$-branes wrapping a holomorphic $H_2$ in a Calabi-Yau
three-fold. In the $D=5$ solutions two of the three gauge-fields are
equal and one of the two scalar fields vanish, which means
\cite{Cvetic:1999xp} that these solutions can be recast as solutions
of Romans' $D=5$ $SU(2)\times U(1)$ gauge theory, with vanishing
abelian gauge-field and with the non-abelian gauge-fields lying in
an abelian subgroup. In our conventions, the non-trivial fields are
given by
\bea
ds^2_5&=&\frac{e^{2f}}{m^2}[ds^2(\bbR^{1,1})+dr^2]+\frac{e^{2g}}{m^2}ds^2(H_2)\nn
X&=&e^{-\varphi}\nn F^3&=&-\frac{1}{{\sqrt 2}m}vol(H_2) \eea
and $f,g,\varphi$ are functions of $r$ that satisfy the differential
equations \bea\label{mnode} e^{-f}\dot
f&=&-\frac{2}{3}e^{-\varphi}-\frac{1}{3}e^{2\varphi}-\frac{1}{6}e^{\varphi-2g}\nn
e^{-f}\dot
g&=&-\frac{2}{3}e^{-\varphi}-\frac{1}{3}e^{2\varphi}+\frac{1}{3}e^{\varphi-2g}\nn
e^{-f}\dot
\varphi&=&-\frac{2}{3}e^{-\varphi}+\frac{2}{3}e^{2\varphi}-\frac{1}{6}e^{\varphi-2g}
\eea
These equations were partially integrated in equation (17) of
\cite{mn}. In \cite{mn} it was shown that there is a solution
describing a flow across dimension from a locally $AdS_5$ region
\be\label{armn}
ds^2_5=\frac{1}{m^2r^2}[ds^2(\bbR^{1,1})+ds^2(H^2)+dr^2] \ee down to
an exact $AdS_3\times H_2$ solution given by \bea\label{gssmn}
ds^2_5&=&\frac{1}{m^22^{4/3}}\left[ds^2(AdS_3)+ ds^2(H_2)\right]\nn
X&=&2^{-1/3}\nn F^3&=&-\frac{1}{{\sqrt 2}m}vol(H_2) \eea

We can now uplift this $AdS_3\times H_2$ solution to $D=11$ using \reef{ul}.
The solution we obtain is particularly simple because $\bar \Delta=1$ and
it is precisely the solution first found in \cite{gk}
which describes an M5-brane wrapping an $H_2\times H_2$ embedded in
a product of two Calabi-Yau two-folds. Moreover, the flow across dimension
solution uplifts to a subclass of flow solutions studied
in \cite{gk}. In particular if one sets $e^{-\varphi}=2^{-1/3}e^{-5\lambda_2/3}$,
$e^{2f}=2^{-4/3}e^{-2\lambda_2/3}e^{2\bar f}$,
$e^{-2g}=2^{4/3}e^{2\lambda_2/3}e^{-2g_2}$ and substitutes into
the differential equations given in \reef{mnode},
one obtains the differential equations in (3.5) of
\cite{gk} after restricting to the case
that $e^{2\lambda_1}=e^{-3\lambda_2}$ and $e^{2g_1}=e^{-\lambda_2}$.

\subsection{Uplifting the Klemm-Sabra magnetic string solution}
As we have already noted, any solution of minimal $D=5$ gauged
supergravity is also a solution of Romans' $D=5$ theory. Consider the
supersymmetric magnetic string solution of \cite{ks}, which we can
write \bea\label{pig}
 ds^2&=&r^{1/2}\left(\frac{r}{3}-\frac{1}{r}\right)^{3/2}
ds^2(\bbR^{1,1})
+\frac{1}{9m^2}\left(\frac{r}{3}-\frac{1}{r}\right)^{-2} dr^2
+\frac{1}{9m^2}r^2ds^2(H_2)\nn G&=&\frac{1}{\sqrt
2}F^3=-\frac{1}{3m}vol(H_2)\nn X&=&1 \eea This interpolates from an
asymptotic locally $AdS_5$ region, with spatial slices
$\bbR^{1,1}\times H_2$ in Poincar\'e coordinates, to an $AdS_3\times
H_2$ solution, which can be written \bea\label{pigt}
ds^2&=&\frac{4}{9m^2}[ds^2(AdS_3)+\frac{3}{4}ds^2(H_2)]\nn
G&=&\frac{1}{\sqrt 2}F^3=-\frac{1}{3m}vol(H_2)\nn X&=&1 \eea

These solutions can be uplifted on\footnote{Using the results of
\cite{Buchel:2006gb} we can uplift on an arbitrary five-dimensional Sasaki-Einstein space; the resulting
$AdS_3$ solutions have already been presented in \cite{Gauntlett:2006qw}.}
an $S^5$ to type IIB supergravity using the formulae in \cite{Cvetic:1999xp}.
The uplifted solutions are dual to D3-branes wrapping a holomorphic $H_2$ embedded in a $CY_4$ (see \cite{naka}).
In particular the uplifted solution \reef{pig} describes the flow across dimension from the
$AdS_5$ region down to the $AdS_3$ fixed point, which is dual to a $(0,2)$ SCFT.

The solution \reef{pig} can also be uplifted to $D=11$ using \reef{ul}. It
then describes a flow across dimension of the $d=4$ SCFT living on
M5-branes wrapped on $H_2$ in $CY_2$, placed on $\bbR^{1,1}\times
H_2$, down to a $d=2$ $(0,2)$ SCFT. As far as we can tell this is a
new solution of $D=11$ supergravity.

We also note, as somewhat of an aside, that the Klemm-Sabra solution can also be
uplifted to $D=11$ in another way. First recall the $N=1$
$AdS_5\times_w {\cal N}_6$ solution of $D=11$ supergravity which
describes the $N=1$ $d=4$ CFT arising on M5-branes wrapping a
holomorphic $H_2$ in a $CY_3$ \cite{mn}. The consistent KK reduction
on this ${\cal N}_6$ down to minimal $D=5$ gauged supergravity was
carried out in \cite{gov}. The solution \reef{pig}, thought of as a
solution of minimal gauged supergravity,
can thus be uplifted on ${\cal N}_6$. The
resulting $D=11$ solution describes the flow across dimension of the
$d=4$ CFT placed on $\bbR^{1,1}\times H_2$ down to the $d=2$ (0,2) CFT
which is dual to the uplifted $AdS_3\times H_2$ solution \reef{pigt}. Moreover,
one finds that the uplifted flow solution and the uplifted $AdS_3$
solution are identical to the corresponding $D=11$ solutions that
describe M5-branes wrapping a holomorphic $H_2\times H_2$ in a
$CY_4$ which were found in \cite{gkw}.

This story can be generalised further by noting that the solution of
\cite{mn} is just one example of several infinite classes of
explicit $AdS_5\times {\cal N}_6$ solutions of $D=11$ supergravity
that were found in \cite{gmsw}, all of which are dual to $N=1$ $d=4$
CFTs. The results of \cite{gov} allow us to uplift the Klemm-Sabra
solution \cite{ks} on any of these ${\cal N}_6$. The resulting
$D=11$ solutions are dual to the flow across dimension of the $d=4$
CFTs on $\bbR^{1,1}\times H_2$ down to $d=2$ (0,2) CFTs which are
dual to the uplifted $AdS_3\times H_2$ solutions. In particular, if
one uplifts the $AdS_3\times H_2$ solution on these ${\cal N}_6$ one
finds solutions that should be included in the general constructions
of \cite{Gauntlett:2006qw}, but we have not checked this in detail.

\subsection{Uplifting Romans' magnetovac solutions}
The solutions of Romans' theory that we discussed in the previous
two subsections are in fact special cases of a more general class of
supersymmetric magnetovac solutions on $AdS_3\times S^2$, $T^2$ and
$H_2$ that were first constructed (earlier) by Romans in \cite{rom}.
The non-abelian $SU(2)$ gauge fields lie in an abelian subgroup and
in addition the $U(1)$ gauge field is, in general, also active.
Specifically, the solutions, which are parametrised by the positive
constant $x$, can be written \bea\label{mvs}
ds^2_5&=&\frac{4x^{4/3}}{m^2(2x+1)^2}\left[ds^2(AdS_3)+ \rtw
ds^2(\Sigma_l)\right]\nn G&=&-\frac{4(2x-1)}{m(2x+1)^2}\rtw
vol(\Sigma_l)\nn F^3&=&-\frac{8x}{{\sqrt 2}m(2x+1)^2}\rtw
vol(\Sigma_l)\nn X&=&x^{1/3} \eea with $l=0,\pm 1$ and $\Sigma_l$ is
$T^2$, $S^2$, $H^2$, respectively.
%The potential $\dub$ is defined so that $d\dub=Vol(\Sigma_l)$.
This is a supersymmetric solution provided that
\bea
l=\frac{4(1-4x)}{(2x+1)^2}\rtw
\eea
In particular, when $0<x<1/4$ we take $l=1$, when $x=1/4$ we take $l=0$ and
when $x>1/4$ we take $l=-1$. When $x=1/4$ we can set $\rtw=1$ after scaling the $T^2$.

Note that when $x=1/2$ the $U(1)$ gauge field vanishes and the corresponding
$AdS_3\times H_2$ solution is precisely the $AdS_3\times H_2$ solution \reef{gssmn} that we discussed above.
This solution actually preserves twice as much supersymmetry as the generic solution.
On the other hand when $x=1$ we get the $AdS_3\times H_2$ solution of minimal gauged supergravity that
we presented in \reef{pigt}.

The general magnetovacs can all be uplifted to $D=11$ using
\reef{ul} to obtain new supersymmetric solutions of $D=11$
supergravity. It would be interesting to study these solutions
further.
%It seems plausible that they are in the class
%of explicit solutions found in \cite{Gauntlett:2006qw}.
It would be
interesting to see if the solutions lie in the class of $AdS_3$ solutions that arise from
the ``K\"ahler-4'' class of Minkowski solutions that were discussed in \cite{Figueras:2007cn}.

The general magnetovac solutions \reef{mvs} can also be uplifted to type IIB on an $S^5$ using
the results of \cite{Lu:1999bw}. We present the explicit results in the
appendix where we also carry out an independent check of the preservation of supersymmetry
using the results of \cite{nakwoo}.

\section{Final comments}
Through inspired guesswork we have constructed the non-linear KK ansatz,
at the level of bosonic fields, for the reduction of $D=11$ supergravity to
Romans' $D=5$ gauged supergravity using the most general $AdS_5\times_w{\cal N}_6$
solutions of $D=11$ supergravity that are dual to $N=2$ SCFTs in $d=4$.
Invoking the argument of \cite{Cvetic:2000dm}
we can conclude that it should be possible
to extend this result at the level of the fermions,
though doing this explicitly would be very difficult.
A less ambitious goal would be to show that for the bosonic configurations
that we are considering, a supersymmetric solution of the Romans' theory
uplifts to a supersymmetric solution of $D=11$ supergravity. This type
of result was shown for other cases in \cite{Buchel:2006gb,gov,gv}.

Another extension of this work would be to show that for the most
general $AdS_5\times_w{\cal N}_5$ solutions of type IIB supergravity
that are dual to $N=2$ $d=4$ SCFTs, there is also a consistent KK
reduction to Romans' theory. However, before this can be
investigated, using the techniques of this paper, the classification
of such solutions, refining the results of \cite{Gauntlett:2005ww},
needs to be carried out.

There is now substantial evidence that the conjecture of \cite{gv}
concerning consistency of KK truncations is correct, having been
verified in several cases. It would be nice to have a rigorous
supergravity proof (perhaps building on the work of
\cite{Pope:1987ad}) independent of a case by case construction.
Ideally, such an analysis would provide an algorithmic prescription
for constructing the non-linear KK ansatz, which, so far, has been
found essentially by trial and error. It would also be nice to have
a general proof from the dual SCFT point of view and some discussion
in this direction has appeared in \cite{David:2007ak}.

\section*{Acknowledgements}
We would like to thank Mirjam Cveti\v{c}, Mike Duff, Oisin Mac Conamhna
and Peter van Nieuwenhuizen for discussions. OV would like to thank the Simons
Workshop 2007, Stony Brook, where part of this work was done. JPG is
supported by an EPSRC Senior Fellowship and a Royal Society Wolfson
Award. OV is supported by the Spanish Ministry of Science and
Education through a postdoctoral fellowship and partially through
the research grant FIS2005-02761.

\appendix

\section{Consistency of the KK ansatz} \label{KKdetails}

In this appendix we provide some details of the proof that the KK
reduction ansatz (\ref{defmet}), (\ref{ansG}) is indeed consistent,
i.e., that it satisfies the equations of motion of $D=11$
supergravity, provided that the field equations of Romans' $D=5$
$SU(2) \times U(1)$ gauged supergravity are imposed.

Our conventions for $D=11$ supergravity follow those of
\cite{Gauntlett:2002fz}. In particular, the equations of motion are
given by
\bea \label{Bianchi11} \dd G_4&=&0 \ , \\ \label{G4eom11} \dd
*_{11}G_4&=&-\tfrac{1}{2}G_4\wedge G_4 \ ,\\ \label{Einstein11}
R_{AB}&=&T_{AB} \ , \eea
where we have defined
\be \label{rhsEinstein}
T_{AB}=\tfrac{1}{12}G_{4AC_1C_2C_3}G_{4B}{}^{C_1C_2C_3}
-\tfrac{1}{144}g_{AB}G_{4C_1C_2C_3C_4}G_4^{C_1C_2C_3C_4} \ . \ee

The frame of the deformed metric \reef{defmet} is taken to be
\bea \bar e^\mu&=&\l^{-1/2}X^{-1/6}\Delta^{1/6}e^\mu\nn \bar
e^1&=&X^{1/3}\Delta^{1/6}e^1\nn \bar e^2&=&X^{1/3}\Delta^{1/6}e^2\nn
\bar e^3&=&X^{5/6}\Delta^{-1/3}\hat e^3\nn \bar
e^4&=&X^{1/3}\Delta^{1/6}e^4\nn \bar
e^a&=&X^{-2/3}\Delta^{-1/3}\frac{\l\r}{2m}(f^a+{\sqrt 2}mk^a_iA^i)
\label{defframe} \eea
where $e^\mu$, $\mu=0, \ldots,4$, is a frame for the
five-dimensional metric $\dd s^2_5$, $(e^1, \ldots , e^6)$ is the
orthonormal frame for the internal undeformed space ${\cal N}_6$
introduced in subsection \ref{undefgeom}, $\hat{e}^3$ is given in
(\ref{shift}), $f^a$, $a=5,6$, is a frame for the unit two sphere
and $k^a_i$ are the components of the three Killing vectors on this
two-sphere with respect to the frame $f^a$. These Killing vectors
satisfy the $SU(2)$ Lie algebra
\be [k_i,k_j]=\e_{ijk}k_k \ee
and also
\be\label{kid} k^{ia} k^j_a=\delta^{ij}-\mu^i\mu^j,\qquad
k^a_ik^b_i=\delta^{ab} \ . \ee
It is also useful to rewrite $D \mu^i$ in (\ref{shift}) as
%\be
%\e^{ab}k^b_i\epsilon^{ijk}\mu^j=k^a_k \ee and that
\be D\mu^i=\e^{ab}k_b^i(f_a+{\sqrt 2}m k_a^k A^k) \ee
and to note that
\begin{equation} \label{DDmu}
DD \mu^i = \sqrt{2}m \epsilon_{ijk}F^k\mu^j =\sqrt{2}m \epsilon^{ab} k^i_b k^k_a F^k \; .
\end{equation}

\subsection{The four-form equations}

The KK ansatz (\ref{ansG}) for the four-form satisfies the Bianchi
identity (\ref{Bianchi11}) provided the $D=5$ gauge fields satisfy
the Bianchi identities that can be obtained by differentiating
(\ref{Bianchis}), and that the field equation (\ref{eomromans3}) for
the $SU(2)$ field strength is imposed. In order to verify this one
needs to use the relations (\ref{65}) and that the differentials of
$\hat{e}^3$ and $D \mu^i$ give field strength contributions, as in
(\ref{DDmu}).
%({\it e.g.}, $DD\mu^i = \sqrt{2}m \epsilon^{ijk} \mu^j F^k$).

To check that the KK ansatz (\ref{defmet}), (\ref{ansG}) satisfies
the $D=11$ four-form equation of motion (\ref{G4eom11}) is somewhat
more involved. Imposing the $D=5$ field equation (\ref{eomromans4})
for simplicity, the Hodge dual, with respect to the deformed metric
(\ref{defmet}), of (\ref{ansG}) reads
{\setlength\arraycolsep{2pt}
\begin{eqnarray} \label{starG}
%\!\!\!\!\!
*_{11} G_4 &=&  *_{11} \tilde{G}_4 \ + X^4
*G \wedge
\lambda^{1/2} \rho \ \hat{e}^{1234} \nonumber \\
&&  +  \tfrac{1}{2\sqrt{2}m}  X^{-2}
*F^i \wedge \Big[\rho
\sqrt{1-z} \epsilon_{ijk} \mu^j D\mu^k \wedge e^{124}
 \nonumber \\ &&  \qquad \qquad
-\tfrac{1}{4m} z  \mu_i \epsilon_{hjk} \mu^h D \mu^j \wedge D\mu^k
\wedge [ \lambda^{-2} e^{12} + \lambda^{-1/2} \rho X \Delta^{-1}
\hat{e}^{34} ] \Big]
 \nonumber \\
&&   + \tfrac{1}{2\sqrt{2}m}  F^i \wedge \rho \ \hat{e}^{123} \wedge
\Big[ \lambda^{1/2}  \ \epsilon_{ijk} \mu^j D\mu^k \wedge e^4
\nonumber \\ &&  \qquad \qquad  - \tfrac{1}{4m} X^{-2} \Delta^{-1}
\sqrt{z(1-z)} \ \mu_i
\epsilon_{hjk} \mu^h D \mu^j \wedge D\mu^k  \Big] \nonumber \\
&&  + \Big[ \tfrac{i}{16\sqrt{2}m^3} F \wedge z \ \epsilon_{ijk}
\mu^i D\mu^j \wedge  D \mu^k \wedge (e^1-ie^2) \wedge (\lambda^{-2}
e^4 +i X\Delta^{-1} \lambda^{-1/2} \rho \hat{e}^3 )
\nonumber \\
&&  -\tfrac{1}{16\sqrt{2}m^2} C \wedge X^{-2} \Delta^{-1} \rho
\sqrt{z(1-z)} \ \epsilon_{ijk} \mu^i D\mu^j \wedge  D \mu^k \wedge
(e^1-ie^2) \wedge \hat{e}^{34} %\nonumber \\ &&
+ c.c. \Big] \; , \nonumber \\
\end{eqnarray}
}%where the $D=5$ field equation (\ref{eomromans4}) has been imposed for simplicity, and
where $*_{11} \tilde{G}_4$ is the Hodge dual, with respect to the
metric (\ref{defmet}), of (\ref{ansGtilde}), namely,
\begin{eqnarray} \label{startildeG}
  *_{11} \tilde{G}_4 &=& 3 X^{-1} * \dd X \wedge \rho \sqrt{1-z}
e^{124} \nonumber \\
&& \frac{1}{\lambda^{9/2} \rho^2} *{\oneone} \wedge
\Big\{\frac{3\lambda \rho^2}{\sqrt{1-z}} X^{-1}
*_6\big[[(\dd\l)_4 e^{12}
         - (\dd\l)_2 e^{14}
         + (\dd\l)_1 e^{24}]
\wedge e^{56}\big] \nonumber \\
&& + X^{-2} \frac{\lambda^{1/2}}{\sqrt{1-z}}
*_6 \big[ [\Delta \dd(z\rho) + \rho (1-z)(X-X^{-2}) \dd z ] \wedge e^{356}
\big] \nonumber \\
&& -2m (X -X^{-2}) z \big[ \lambda \rho X e^{34} + \lambda^{-1/2}
\Delta e^{12} \big] \Big\} \; .
\end{eqnarray}
Here, $*_6$ is the Hodge dual with respect to the undeformed metric
$ds^2 ({\cal N}_6)$ in (\ref{undef11metric}). In fact, the presence
in (\ref{startildeG}) of the volume form $* {\oneone}$, corresponding
to the spacetime metric $\dd s_5^2$, allows one to write
(\ref{startildeG}) in terms of the frame $(e^1, \ldots , e^6)$ of
the undeformed metric $ds^2 ({\cal N}_6)$, once the contributions
from the scalar field $X$ have been taken into account.

Computing the exterior derivative of (\ref{startildeG})
we find
\begin{eqnarray} \label{dstartildeG}
\dd *_{11} \tilde{G}_4 &=& \big[ 3 \dd ( X^{-1} * \dd X) +4m^2 (X^2
-X^{-1}) *{\oneone} \big] \wedge \rho \sqrt{1-z} \ e^{124} \; ,
\end{eqnarray}
where we used the field equation for the undeformed four-form
(\ref{6.9}), which can be written
\begin{eqnarray}
&& \dd \Big\{ \frac{1}{\lambda^{9/2} \rho^2 \sqrt{1-z}} \Big[
\lambda^{1/2}
*_6
 [ \dd(z\rho) \wedge e^{356} ] \nonumber \\ && \qquad
 + 3 \lambda \rho^2 *_6\big[[(\dd\l)_4 e^{12}
         - (\dd\l)_2 e^{14}
         + (\dd\l)_1 e^{24}]
\wedge e^{56}\big] \Big] \Big \} =0 \; .
\end{eqnarray}
Next, differentiating (\ref{starG}) with the help of
(\ref{dstartildeG}) and (\ref{65}),
%taking again into account that
%the differentials of $\hat{e}^3$ and $D \mu^i$ give field strength
%contributions, and
and wedging (\ref{ansG}) with itself, one can compute
{\setlength\arraycolsep{0pt}
\begin{eqnarray}
%\ker[-30em]
\dd *_{11} G_4 && +\tfrac{1}{2} G_4 \wedge G_4 = \nonumber \\[10pt]
&&  \Big[ 3 \dd(X^{-1} *\dd X) -X^4 *G \wedge G +\tfrac{1}{2}
X^{-2} (*F^i \wedge F^i + *C \wedge \bar{C}) \nonumber \\
&& \qquad \qquad \qquad \qquad +4m^2 (X^2-X^{-1}) \,
{*\oneone} \Big] \wedge \rho \sqrt{1-z} e^{124} \nonumber \\
&&   + \Big[ d(X^4\, {*G}) + \tfrac{1}{2} F^i\wedge F^i +
\tfrac{1}{2} {\bar C}\wedge C \Big] \wedge \lambda^{1/2} \rho \
\hat{e}^{1234}
\nonumber \\
&&  + \tfrac{1}{2\sqrt{2}m}  \Big[ D(X^{-2}\, {*F^i}) + F^i\wedge G
\Big] \wedge \Big[\rho \sqrt{1-z} \epsilon_{ijk} \mu^j D\mu^k \wedge
e^{124} \nonumber \\
&& \qquad \qquad -\tfrac{1}{4m} z  \mu_i \epsilon_{hjk} \mu^h D
\mu^j \wedge D\mu^k \wedge [ \lambda^{-2} e^{12} + \lambda^{-1/2}
\rho X \Delta^{-1} \hat{e}^{34} ] \Big] .
\end{eqnarray}
}This indeed shows that the KK ansatz (\ref{defmet}), (\ref{ansG})
satisfies the $D=11$ four-form equation of motion (\ref{G4eom11}),
provided that the five-dimensional fields satisfy the field
equations (\ref{eomromans1})--(\ref{eomromans4}) of Romans' $D=5$
$SU(2) \times U(1)$ gauged supergravity.

\subsection{The Einstein equations}

In order to check the $D=11$ Einstein equations (\ref{Einstein11}),
we first give explicit expressions for the tensor $T_{AB}$,
that defines the right hand side, in terms of the frame
\reef{defframe}. Substituting the expression \reef{ansG} of the KK
ansatz for $G_4$ into (\ref{rhsEinstein}) we find, for the external
components,
\bea\label{tmn} &&\l^{-1}X^{-1/3}\Delta^{1/3}T_{\mu\nu}=\nn
&&\frac{zX^5}{2\Delta}(G_{\mu\rho}G_{\nu}{}^\rho-\frac{1}{6}\eta_{\mu\nu}G_{\rho\sigma}G^{\rho\sigma})
-\frac{1}{24}(
(1-z)X^{-4}\Delta^{-1}+2X^{-2})\eta_{\mu\nu}F^i_{\r\s}F^{i\r\s}\nn
&&+\frac{1}{4}(X^{-2}+(1-z)X^{-4}\Delta^{-1})F^i_{\mu\r}F^i_{\nu}{}^\r
+\frac{1}{4}X^{-1}z\Delta^{-1}F^i_{\mu\r}F^j_{\nu}{}^\r\mu^i\mu^j
\nn &&\tfrac{1}{2}X^{-2}[C_{(\mu}{}^\r\bar
C_{\nu)\r}-\frac{1}{6}\eta_{\mu\nu}C_{\r\s} \bar C^{\r\s}]
-\tfrac{1}{24}(1-z)X^{-4}\Delta^{-1}\eta_{\mu\nu}C_{\r\s} \bar
C^{\r\s}\nn &&+\frac{3}{2}(1-z)z\Delta^{-2}X^{-3}[3\nabla_\mu
X\nabla_\nu X-\eta_{\mu\nu}(\nabla X)^2]
-\frac{3}{2}\frac{\Delta+zX}{\Delta^2 X^3
(1-z)\lambda^3}(\nabla\lambda)^2\nn
&&-\frac{2m\rho(X(2+z)+X^{-2}(1-z))}{{\sqrt{1-z}}\Delta^2
X^3}(d\lambda)_4\nn &&-\frac{2m^2}{3\Delta^2
X^5}[2z^2X^9+z(7-5z)X^6+(9-8z+4z^2)X^3+z(1-z)] \eea
where we note, for example, that $G_{\mu\nu}$ are the components of
$G$ with respect to the $D=5$ frame $e^\mu$.
For the mixed components we find
 {\setlength\arraycolsep{2pt}
\begin{eqnarray}
T_{\mu 1} &=& \tfrac{9}{2}  X^{-13/6} \Delta^{-7/3} \lambda^{-1/2} z
\
(\dd \lambda)_1 \ \nabla_\mu X \; , \nonumber \\[10pt]
T_{\mu 2} &=& \tfrac{9}{2}  X^{-13/6} \Delta^{-7/3} \lambda^{-1/2} z
\
(\dd \lambda)_2 \ \nabla_\mu X \; , \nonumber \\[10pt]
T_{\mu 3} &=& X^{4/3} \Delta^{-5/6} \lambda \sqrt{1-z} \ \big[
-\tfrac{3}{2}  z \Delta^{-1} \ G_{\mu \nu} \nabla^\nu X %\nonumber \\ &&
+ \tfrac{1}{16} X^{-4} \epsilon_{\mu \lambda \nu \rho \sigma} ( F^{i
\lambda \nu} F^{i \rho \sigma} + C^{ \lambda \nu} \bar{C}^{ \rho
\sigma}) \big] , \nonumber \\[10pt]
T_{\mu 4} &=&   X^{-7/6} \Delta^{-7/3} \lambda^{-1/2} z \left[
\tfrac{9}{2} X^{-1} (\dd \lambda)_4 + 3m \rho^{-1} \sqrt{1-z} \ [X^2
z + X^{-1}(3-z)] \right]
\nabla_\mu X , \nonumber \\[10pt]
T_{\mu a} &=& \tfrac{1}{2\sqrt{2}} X^{-1/6} \Delta^{-5/6}
\lambda^{5/2}\rho \ k^i_a \big[ {3} (1-z) X^{-3} \Delta^{-1} \
F^i_{\mu \nu}
\nabla^\nu X + \tfrac{1}{4} X^2 \epsilon_{\mu \lambda \nu \rho \sigma} F^{ \lambda \nu}G^{ \rho \sigma} \big] % \nonumber \\[10pt]
\end{eqnarray}
}Finally, to write the internal components of $T_{AB}$ it proves
convenient to introduce, for $n=1,2,4$,
{\setlength\arraycolsep{2pt}
\begin{eqnarray}
U_n&=& \tfrac{\lambda^{1/2}}{\sqrt{1-z}} X^{1/6} \Delta^{-7/6} \big[
3 \rho X^{-1} (\dd \lambda)_n + 2m\sqrt{1-z} X [Xz+(3-z)X^{-2} ]
\delta_{n4} \big] \; , \nonumber \\[10pt]
V_n &=& \tfrac{1}{\lambda \sqrt{1-z}} X^{2/3} \Delta^{-2/3} \big[ 3
X^{-2} (\dd \lambda)_n -2m \lambda^3 \rho \sqrt{1-z} (X-X^{-2}) \
\delta_{n4} \big] \; ,
\end{eqnarray}
}so that (\ref{ansGtilde}) can be written in the frame
(\ref{defframe}) as
\begin{eqnarray}
\tilde{G}_4 = ( V_4 \bar{e}^{12} + U_1 \bar{e}^{13} - V_2
\bar{e}^{14} + U_2 \bar{e}^{23}  + V_1 \bar{e}^{24} - U_4
\bar{e}^{34} ) \wedge \bar{e}^{56} \, .
\end{eqnarray}
Then one finds, for the non-vanishing internal components ($m,n,p
\in \{ 1,2,4 \}$, $a,b \in \{5,6 \}$):
{\setlength\arraycolsep{2pt}
\begin{eqnarray}
T_{mn} & = & \tfrac{1}{12} X^{4/3} \Delta^{-4/3} \lambda z \big[ -
X^4 G_{\mu \nu} G^{\mu \nu} + \tfrac{1}{2} X^{-2} \left(F^i_{\mu \nu} F^{i
\mu \nu} + C_{\mu \nu} \bar{C}^{ \mu \nu}\right) \big]
\ \delta_{mn} \nonumber
\\ && -\tfrac{3}{2} \lambda z(1-z) \Delta^{-7/3} X^{-8/3} \nabla_\mu X
\nabla^\mu X \ \delta_{mn} \nonumber \\
&& +\tfrac{1}{2} (U_m U_n -V_m V_n) + \tfrac{1}{6} (-U_p U^p +2V_p V^p) \ \delta_{mn} , \nonumber \\[10pt]
T_{13}&=& \tfrac{1}{2} (U_4 V_2 -U_2 V_4)  , \nonumber \\[10pt]
T_{23}&=& \tfrac{1}{2} (U_1 V_4 -U_4 V_1)  , \nonumber \\[10pt]
T_{33} &=& -\tfrac{1}{12} \lambda z  X^{16/3} \Delta^{-4/3} G_{\mu
\nu} G^{\mu \nu} \nonumber \\
& & +\tfrac{1}{24}  \lambda X^{-5/3} \Delta^{-4/3} [Xz + 3X^{-2}
(1-z)] \left( F^i_{\mu \nu} F^{i \mu \nu} + C_{\mu \nu}
\bar{C}^{ \mu \nu}\right)
 \nonumber \\
&& +3 \lambda z(1-z) \Delta^{-7/3} X^{-8/3}  \nabla_\mu X
\nabla^\mu X \nonumber \\
&& + \tfrac{1}{6}  (-U_p U^p +2V_p V^p)
\; , \nonumber \\[10pt]
T_{3a} &=& \tfrac{1}{4\sqrt{2}} X^{5/6} \Delta^{-4/3}
\lambda^{5/2}\rho \sqrt{1-z} \ k^i_a F^i_{\mu \nu} G^{\mu \nu}
\; , \nonumber \\[20pt]
T_{ab} &=&  -\tfrac{1}{6} \lambda z X^{4/3} \Delta^{-4/3}  \big[
-X^4 G_{\mu \nu} G^{\mu \nu} + \tfrac{1}{2} X^{-2} \left(F^i_{\mu \nu}
F^{i \mu \nu} +  C_{\mu \nu} \bar{C}^{ \mu \nu}\right)
\big] \ \delta_{ab} \nonumber
\\ && +\tfrac{1}{8}  X^{-2/3} \Delta^{-4/3} \lambda z k^i_a k^j_b F^i_{\mu \nu} F^{j \mu
\nu} \nonumber \\
&&  +3 \lambda  z(1-z) \Delta^{-7/3} X^{-8/3}  \nabla_\mu X
\nabla^\mu X  \ \delta_{ab}  \nonumber \\
&& + \tfrac{1}{3} (-U_p U^p +2V_p V^p) \ \delta_{ab} \; .
\end{eqnarray}
}

The spin connection corresponding to the deformed metric
(\ref{defmet}) can be computed in the frame (\ref{defframe}) and we
find, for the external components,
{\setlength\arraycolsep{2pt}
\begin{eqnarray}
\bar{\omega}^{\mu \nu} &=& \omega^{\mu \nu} + \lambda^{1/2}
X^{-17/6} \Delta^{-7/6} (1-z) \nabla^{[\mu} X \ \bar{e}^{\nu]}
% \nonumber \\ &&
\ -\tfrac{1}{2} X^{7/6} \Delta^{-2/3} \lambda^{1/2} \sqrt{1-z} \
G^{\mu \nu} \bar{e}^3 \nonumber \\
&& -\tfrac{1}{2\sqrt{2}} X^{-1/3} \Delta^{-2/3} \lambda^2 \rho \
k^i_a F^{i \mu \nu} \bar{e}^a \; ,
\end{eqnarray}
}for the mixed components,
{\setlength\arraycolsep{2pt}
\begin{eqnarray}
\bar{\omega}^{\mu 1} &=&  -\tfrac{1}{2} X^{-7/3} \Delta^{-7/6}
\lambda^{-1}  (\dd \lambda)_1 \ \bar{e}^\mu % \nonumber \\ &&
\ - \tfrac{1}{2} \lambda^{1/2} z   X^{1/6} \Delta^{-7/6}  \nabla^\mu
X \ \bar{e}^1
%\nonumber \\
%&& -X^{-1/3} \Delta^{1/3} (1-X \Delta^{-1}) \tfrac{(1+2z)
%}{4 \lambda \sqrt{1-z}} (\dd \lambda)_1 B^\mu  \bar{e}^3 \nonumber \\
%&& + \tfrac{1}{2\sqrt{2}}  X^{7/6} \Delta^{1/3} (1-X^{-2}
%\Delta^{-1}) \lambda^{1/2} \rho (\dd \lambda)_1 k^i_a A^{i \mu}
%\bar{e}^a
\nonumber \\[10pt]
\bar{\omega}^{\mu 2} &=&  -\tfrac{1}{2} X^{-7/3} \Delta^{-7/6}
\lambda^{-1}  (\dd \lambda)_2 \ \bar{e}^\mu % \nonumber \\ &&
\ -\tfrac{1}{2} \lambda^{1/2} z X^{1/6} \Delta^{-7/6} \nabla^\mu X \
\bar{e}^2
%\nonumber \\
%&& -X^{-1/3} \Delta^{1/3} (1-X \Delta^{-1}) \tfrac{(1+2z)
%}{4 \lambda \sqrt{1-z}} (\dd \lambda)_2 B^\mu  \bar{e}^3 \nonumber \\
%&& + \tfrac{1}{2\sqrt{2}}  X^{7/6} \Delta^{1/3} (1-X^{-2}
%\Delta^{-1}) \lambda^{1/2} \rho (\dd \lambda)_2 k^i_a A^{i \mu}
%\bar{e}^a
\nonumber \\[10pt]
\bar{\omega}^{\mu 3} &=&  -\tfrac{1}{2} X^{7/6} \Delta^{-2/3}
\lambda^{1/2} \sqrt{1-z} \ G^\mu{}_{\nu} \ \bar{e}^\nu  \nonumber \\
&& \ -\tfrac{1}{2} \lambda^{1/2} X^{-5/6} \Delta^{-7/6}  [X z +3
X^{-2}(1-z)] \nabla^\mu X   \ \bar{e}^3
%\nonumber \\
%&& -\Delta^{1/2} (1-X \Delta^{-1}) \tfrac{(1+2z) }{4 \lambda
%\sqrt{1-z}} B^\mu \dd \lambda \ -m X^{-1/3} \Delta^{1/3} (1-X
%\Delta^{-1})  \lambda^2 \rho  B^\mu \bar{e}^4
\nonumber \\[10pt]
\bar{\omega}^{\mu 4} &=&   X^{-1/3} \Delta^{-7/6} \lambda^{-1} \big[
 - \tfrac{1}{2} X^{-2}( \dd \lambda)_4
+ \tfrac{2mz}{3\rho} \sqrt{1-z} (X - X^{-2}) \big] \ \bar{e}^\mu
\nonumber \\ &&
 -\tfrac{1}{2} \lambda^{1/2} z X^{1/6} \Delta^{-7/6}  \nabla^\mu X \ \bar{e}^4
%\nonumber \\
%&& -X^{-1/3} \Delta^{1/3} (1-X \Delta^{-1}) \big[ \tfrac{(1+2z)
%}{4 \lambda \sqrt{1-z}} (\dd \lambda)_4 + m \lambda^2 \rho \big] B^\mu  \bar{e}^3 \nonumber \\
%&& +\tfrac{1}{2\sqrt{2}}  X^{7/6} \Delta^{1/3} (1-X^{-2}
%\Delta^{-1}) \lambda^{1/2} \big[ \rho (\dd \lambda)_4 + 2m
%\sqrt{1-z} \big] k^i_a A^{i \mu} \bar{e}^a
\nonumber  \\[10pt]
\bar{\omega}^{\mu a} &=&   -\tfrac{1}{2\sqrt{2}} X^{-1/3}
\Delta^{-2/3}
\lambda^2 \rho  k^{ai} F^{i \mu}{}_{\nu} \  \bar{e}^\nu % \nonumber \\ &&
\  + \lambda^{1/2} z X^{1/6} \Delta^{-7/6}  \nabla^\mu X  \
\bar{e}^a
% \nonumber \\ &&
% + \tfrac{1}{2\sqrt{2}} X^{3/2} \Delta^{1/2} (1-X^{-2} \Delta^{-1})
%\lambda^{1/2} \rho \ k^{ai} A^{i \mu} \dd \lambda \nonumber \\
%&& +\tfrac{m}{\sqrt{2}} X^{7/6} \Delta^{1/3} (1-X^{-2} \Delta^{-1})
%\lambda^{1/2} \sqrt{1-z} \ k^{ai} A^{i \mu} \bar{e}^4 \; ,
\end{eqnarray}
}and for the non-vanishing internal components,
{\setlength\arraycolsep{2pt}
\begin{eqnarray} \label{omegaintdef}
\bar{\omega}^{12} &=&
%-   [ \tfrac{3\rho}{2\sqrt{1-z}} (\dd \lambda)_4 + 2m  ] \ B +
-m B + M_2 \bar{e}^1   -M_1 \bar{e}^2  -N_4 \bar{e}^3 \nonumber  \\%[10pt]
\bar{\omega}^{13} &=&
%- X^{-1/2} \Delta^{1/2} (1+X \Delta^{-1}) \tfrac{\rho(1+2z)}{4 \sqrt{z(1-z)}} (\dd \lambda)_1 \ B +
P_4 \bar{e}^2  -Q_1 \bar{e}^3   -P_2 \bar{e}^4   \nonumber  \\%[10pt]
\bar{\omega}^{14} &=&
%\tfrac{3\rho}{2\sqrt{1-z}} (\dd \lambda)_2  \ B  +
M_4 \bar{e}^1 +N_2 \bar{e}^3  -M_1 \bar{e}^4   \nonumber  \\%[10pt]
\bar{\omega}^{1a} &=&
% \tfrac{1}{2\sqrt{2}} X \Delta^{1/2} (1+X^{-2} \Delta^{-1}) \rho (\dd \lambda)_1 k^{ia} A^i
- R_1 \bar{e}^a \nonumber  \\%[10pt]
\bar{\omega}^{23} &=&
% -X^{1/2} \Delta^{1/2} (1+X \Delta^{-1}) \tfrac{\rho(1+2z)}{4 \sqrt{z(1-z)}} (\dd \lambda)_2 \ B
-P_4 \bar{e}^1  -Q_2 \bar{e}^3   +P_1 \bar{e}^4   \nonumber  \\%[10pt]
\bar{\omega}^{24} &=&
% -\tfrac{3\rho}{2\sqrt{1-z}} (\dd \lambda)_1  \ B +
M_4 \bar{e}^2 -N_1 \bar{e}^3  -M_2 \bar{e}^4   \nonumber  \\%[10pt]
\bar{\omega}^{2a} &=&
%  \tfrac{1}{2\sqrt{2}}   X \Delta^{1/2} (1+X^{-2} \Delta^{-1}) \rho (\dd \lambda)_2 k^{ia} A^i
- R_2 \bar{e}^a \nonumber  \\%[10pt]
\bar{\omega}^{34} &=&
% - X^{-1/2} \Delta^{1/2} (1+X \Delta^{-1}) \big[ \tfrac{\rho(1+2z)}{4 \sqrt{z(1-z)}} (\dd \lambda)_4 + m \sqrt{z} \big] \ B
-P_2 \bar{e}^1  +P_1 \bar{e}^2 +Q_4 \bar{e}^3     \nonumber  \\%[10pt]
\bar{\omega}^{4a} &=&
% \tfrac{1}{2\sqrt{2}}   X \Delta^{1/2} (1+X^{-2} \Delta^{-1}) \lambda^{1/2} \big[ \rho (\dd \lambda)_4 +2m \sqrt{1-z} \big] k^{ia} A^i
- R_4 \bar{e}^a \nonumber  \\%[10pt]
\bar{\omega}^{56} &=& \mu^i A^i - \frac{\mu^3}{\sqrt{(\mu^1)^2 +
(\mu^2)^2}} f^6
\end{eqnarray}
}
where $f^a$, $a=5,6$, was introduced in (\ref{defframe}) and we have
defined, for $n=1,2,4$:
\begin{eqnarray}
M_n &=& \tfrac{1}{6\lambda (1-z)} X^{-1/3} \Delta^{-7/6}  \Big[
\left[9Xz + 6X^{-2} (1-z) \right] (\dd \lambda)_n  \nonumber \\
&&  \quad \qquad \qquad  \qquad \quad + 2m\lambda^3\rho \sqrt{1-z}
\left[X(2+z) +X^{-2} (1-z)\right] \delta_{n4} \Big] ,
\nonumber \\[10pt]
 N_n &=& -\tfrac{\lambda^{1/2}
}{2(1-z)} X^{-5/6} \Delta^{-2/3}  \Big[  3\rho X (\dd \lambda)_n +
2m \sqrt{1-z} \left[ X(1+z) +X^{-2} (1-z) \right] \delta_{n4} \Big]
,
\nonumber \\[10pt]
 P_n &=& \tfrac{\lambda^{1/2} }{2(1-z)} X^{1/6} \Delta^{-2/3}
 \Big[ 3\rho  (\dd \lambda)_n +
2m\sqrt{1-z} \ \delta_{n4} \Big] , \nonumber \\[10pt]
Q_n &=& -\tfrac{1}{6\lambda (1-z)} X^{-1/3} \Delta^{-7/6} \Big[
\left[ 9Xz + 3 X^{-2} (1-z) \right] (\dd \lambda)_n \nonumber \\ &&
\quad \qquad \qquad \qquad \qquad + 4m\lambda^3\rho \sqrt{1-z}
\left[ X(2+z) +X^{-2} (1-z) \right] \delta_{n4} \Big] ,
\nonumber \\[10pt]
R_n &=& X^{-1/3} \Delta^{-7/6} \Big[ \lambda^{-1} X^{-2}  (\dd
\lambda)_n + \tfrac{2m}{3\lambda \rho} \sqrt{1-z} \left[ Xz +X^{-2}
(3-z) \right]
 \delta_{n4} \Big] .
\end{eqnarray}
Notice that, when $X=1$, $B=A^i=0$, the spin connection reduces to
that of the undeformed metric (\ref{undef11metric}). In particular,
the internal components (\ref{omegaintdef}) reduce to the spin
connection of the undeformed $ds^2 ({\cal N}_6)$, that can be
calculated from the equations  (\ref{65}).

The Ricci tensor corresponding to the deformed metric (\ref{defmet})
can now be calculated in the frame  (\ref{defframe}) and, for
illustration, we just record here the expression for its external
components $\bar{R}_{\mu \nu}$. To do this it is convenient to
notice that, for any of the solutions $AdS_5 \times_w {\cal N}_6$
described in subsection \ref{undefgeom}, one has
\bea \nabla^2\lambda+4m^2\lambda^2+
\frac{13z-1}{2\lambda(1-z)}(\nabla\lambda)^2 +
\frac{12mz}{\lambda\rho{\sqrt{1-z}}}(\dd\lambda)_4 =0 , \eea
as can be shown using \reef{65}. Defining the tensor
\bea E_{\mu\nu}&=& R_{\mu\nu} -3 X^{-2}\, \nabla_\m X\, \nabla_\n X
+ \tfrac{4}{3}m^2\,(X^2 + 2  X^{-1})\, \eta_{\m\n}\nn & & -
\tfrac{1}{2} X^4 \, [G_\m{}^\r G_{\n \r} -\tfrac{1}{6} \eta_{\m\n}
\, G_{\r\s}G^{\r\s}] - \tfrac{1}{2} X^{-2}\, [F^{i\ \r}_\m
F^{i}_{\n\r}  - \tfrac{1}{6} \eta_{\m\n}\, F^i_{\r\s}F^{i\r\s}]\nn &
& - \tfrac{1}{2} X^{-2}\,  [ C_{(\m}{}^\r\, \bar C_{\n)\r} -
\tfrac{1}{6} \eta_{\m\n}\, C_{\r\s}\bar C^{\r\s}]\,, \eea
and the scalar
\be S=3\nabla_\mu(X^{-1}\nabla^\mu
X)+4m^2(X^2-X^{-1})-\tfrac{1}{2}X^4G_{\mu\nu}G^{\mu\nu}
+\tfrac{1}{4}X^{-2}F^i_{\mu\nu}F^{i\mu\nu}+\tfrac{1}{4}X^{-2}C_{\mu\nu}
\bar C^{\mu\nu} , \ee
a long calculation reveals that
\be \bar R_{\mu\nu}=  \l X^{1/3}\Delta^{-1/3} \left[
E_{\mu\nu}+\eta_{\mu\nu} \frac{(1-z)}{6X^2\Delta}S \right]
+T_{\mu\nu} , \ee
where $T_{\mu\nu}$ is given in \reef{tmn}. This shows that the
external components of the $D=11$ Einstein equations
(\ref{Einstein11}) are satisfied provided $S=0$ and $E_{\m\n}=0$,
which are precisely the scalar (\ref{eomromans1}) and Einstein
equations (\ref{eomromans5}) of Romans' $D=5$ gauged supergravity.

\section{The magnetovac solutions uplifted to type IIB}
After uplifting the magnetovac solutions \reef{mvs} to type IIB using \cite{Lu:1999bw}
we find that the ten-dimensional metric is given by
\bea\label{pent}
 m^2 ds^2_{10} &=& \frac{4x}{(2x+1)^2} \bar\Delta^{1/2} \left[ds^2(AdS_3)+
\rtw ds^2(\Sigma_l)\right] +\bar\Delta^{1/2}
d\xi^2+\frac{\cos^2\xi}{4\bar\Delta^{1/2}}  d\Omega_2 \nn &&
+\frac{\cos^2\xi}{4\bar\Delta^{1/2}} \big[\sigma_3 -\frac{8x\rtw}{(2x+1)^2}
\dub \big]^2 +\frac{x\sin^2\xi}{\bar\Delta^{1/2}} \big[ d\tau
-\frac{4(2x-1)\rtw}{(2x+1)^2}\dub \big]^2  \eea
where $\bar\Delta=\sin^2\xi+x\cos^2\xi$ and the potential $\dub$ is defined
so that $d\dub=vol(\Sigma_l)$.

We can directly check the supersymmetry of this uplifted IIB solution by
recasting it in the general form found in \cite{nakwoo}. In
particular the solutions are dual to SCFTs with (at least) $(0,2)$
supersymmetry. To do this we first employ a coordinate
transformation so that
\bea
\sigma_3&=&\sigma_3'+\frac{2x}{2x+1}dz\nn d\tau&=&\frac{1}{2x+1}dz
\eea
We then find that the metric takes the form
\bea
m^2ds^2_{10}=e^{2A}\left[ds^2(AdS_3)+e^{-4A}ds^2_6+\frac{1}{4}(dz+P)^2\right]
\eea
where
\bea e^{2A}&=&\frac{4x}{(2x+1)^2}\bar\Delta^{1/2}\nn
e^{-4A}ds^2_6&=&\rtw ds^2(\Sigma_l)+\frac{(2x+1)^2}{4x}d\xi^2
\nn && +\frac{(2x+1)^2\cos^2\xi}{16x\bar\Delta}d\Omega_2
+\frac{(2x+1)^2\cos^2\xi\sin^2\xi}{16x\bar\Delta^2}
[\sigma_3'+\frac{16x(x-1)\rtw}{(2x+1)^2}\dub]^2\nn P&=&-\frac{4 \rtw}{(2x+1)\bar \Delta}[
(2x-1)\sin^2\xi+x\cos^2\xi]\dub+\frac{(2x+1)\cos^2\xi}{2\bar\Delta}\sigma_3'
\eea
After some calculation one can show that $ds^2_6$ is K\"ahler, with
K\"ahler form $J_6$ given by
\bea e^{-4A} J_6 &=& \rtw vol(\Sigma_l) +
\frac{(2x+1)^2 \cos^2 \xi}{16x \bar\Delta}  vol(S^2) \nn &&
-\frac{(2x+1)^2\cos\xi\sin\xi}{8x\bar\Delta} d\xi \wedge
[\sigma_3'+\frac{16x(x-1)\rtw}{(2x+1)^2}\dub] \eea
and that $P$ is a Ricci-form potential for this K\"ahler metric. Further calculation
shows that the warp factor satisfies \be R=8e^{-4A} \ee and that $ds^2_6$ satisfies
\be
\nabla^2{R}+R_{ij}R^{ij}-\frac{1}{2}R^2=0 \ee
This verifies that the IIB solutions preserve $(0,2)$ supersymmetry \cite{nakwoo}.

We have already noted that for $x=1$ the solution corresponds to D3-branes wrapped on a $H_2$ in a $CY_4$
and is dual to a $(0,2)$ SCFT. The $x=1/2$ solution corresponds to D3-branes wrapped on a  $H_2$ in a $CY_3$,
and is dual to a $(2,2)$ SCFT. The solutions for generic $x$ are dual to SCFTs with $(0,2)$ SCFT.
It is natural to wonder if they lie within the class of explicit solutions found in \cite{Gauntlett:2006qw}.
If we perform the coordinate transformation
\bea
\sigma_3&=&\sigma_3'-d\tau'\nn d\tau&=&d\tau'
\eea
the metric \reef{pent} can be written
\bea m^2 ds^2_{10} &=& \frac{4x}{(2x+1)^2} \bar\Delta^{1/2} \left[ds^2(AdS_3)+
\rtw ds^2(\Sigma_l)\right] +\bar\Delta^{1/2}
d\xi^2+\frac{\cos^2\xi}{4\bar\Delta^{1/2}}  d\Omega_2 \nn &&
+\frac{\Z}{4 \bar\Delta^{1/2}} \big[ d\tau'-\frac{\cos^2\xi}{\Z}\sigma_3'-
\frac{8x\rtw}{\Z(2x+1)^2}(-\cos^2\xi+2(2x-1)\sin^2\xi)\dub \big]^2  \nn&&
+\frac{x\cos^2\xi\sin^2\xi}{\bar \Delta^{1/2} \Z}
\big[\sigma_3' + l\dub \big]^2
\eea
where $\Z =4x\sin^2\xi+\cos^2\xi$. One can now compare with the solutions
in \cite{Gauntlett:2006qw} (see eq (2.11) of this reference).

%%%%%%%%%%%%%%%%%%%%%%%%%%%%%%%%%

\end{document}